\documentclass[aps,twocolumn,showpacs]{revtex4}
\usepackage{amsmath}
\usepackage{epsfig}

\begin{document}
	
\title{Fully heavy pentaquarks in quark models}
\author{Ye Yan}\email{201002013@njnu.edu.cn}
\author{Yuheng Wu}\email{191002007@njnu.edu.cn}
\author{Xiaohuang Hu}\email{201001002@njnu.edu.cn}
\author{Hongxia Huang}\email{hxhuang@njnu.edu.cn(Corresponding author)}
\author{Jialun Ping}\email{jlping@njnu.edu.cn(Corresponding author)}
\affiliation{Department of Physics, Nanjing Normal University, Nanjing, Jiangsu 210097, P. R. China}

\begin{abstract}
The fully heavy pentaquarks $cccc\bar{c}$ and $bbbb\bar{b}$ are systematically investigated within the chiral quark model and quark delocalization color screening model.
The results are consistent in both two moelds, and the effect of the channel-coupling is crucial for forming a bound state of the fully heavy pentaquark system.
Three possible fully heavy pentaquarks are obtained, which are the $cccc\bar{c}$ state with $J^P = 1/2^-$ and the mass of $7891.9 \sim 7892.7$ MeV, the $bbbb\bar{b}$ state with $J^P=1/2^-$ and the mass of $23810.1 \sim 23813.8$ MeV, and the $bbbb\bar{b}$ state with $J^P = 3/2^-$ and the mass of $23748.2 \sim 23752.3$ MeV. All these fully heavy pentaquark states are worth searching in future experiments.
\end{abstract}
	
	
\maketitle

\setcounter{totalnumber}{5}
	
\section{\label{sec:introduction}Introduction}
Over the past decades, great progress has been made on studying exotic states, among which the fully heavy tetraquark states $(QQ\bar{Q}\bar{Q}, Q=c,b)$ attracted extensive attention, since such states with very large energy can be accessed experimentally and easily distinguished from other states. In 2020, the LHCb collaboration reported their result on the observation of the fully charm state ($cc\bar{c}\bar{c}$). A narrow structure $X(6900)$, matching the lineshape of a resonance and a broad structure next to the $J/\psi J/\psi$ mass threshold was obtained~\cite{Aaij:2020fnh}. Theoretically, this fully charm tetraquark have been investigated in the framework of the QCD sum
rules~\cite{ZGWang}, the potential model~\cite{MSLiu}, the non-relativistic diquark-antidiquark model~\cite{Lundhammar}, the relativized quark model~\cite{QLv}, string junction picture~\cite{MKarliner2}, and the constituent quark model~\cite{GYang,Jin:2020jfc}. Meanwhile, the molecular configurations for the fully charm tetraquark states were proposed by the perturbation QCD~\cite{Albuquerque}, while the compact structure was obtained
in Holography inspired stringy hadron model~\cite{Sonnenschein}. In the recently theoretical investigations, the experimental data and theoretical
importance on the $cc\bar{c}\bar{c}$ tetraquarks were reviewed in the articles~\cite{JRichard,KChao}.

For the dibaryons, although the deuteron is the only stable state composed of two nucleons, there are possible bound or
resonant dibaryons with and without strange quarks. Both the high strangeness dibaryons $N\Omega$ and $\Omega\Omega$ were proposed by quark models~\cite{Goldman2,Zhang}.
The lattice QCD simulations by HAL QCD Collaboration have studied the six-quark systems containing light or strange quarks and confirmed the existence of the $N\Omega$ and $\Omega\Omega$ bound states with nearly physical quark masses ($m_{\pi}\simeq 146$ MeV and $m_{K}\simeq 525$ MeV)~\cite{Lattice1,Lattice2}. The fully heavy dibaryons were also proposed by both the quark models~\cite{Huang:2020bmb} and the lattice QCD method~\cite{Lattice3,Lattice4}, in which the dibaryon with the highest charm number $\Omega_{ccc}\Omega_{ccc}$ was possible to exist.

Regarding to the pentaquarks, the most noteworthy states in recent years are the hidden-charm pentaquarks. In 2015, the LHCb Collaboration reported two hidden-charm pentaquark states $P_{c}(4380)$ and $P_{c}(4450)$ in the $J/\psi p$ invariant mass spectrum of $\Lambda^{0}_{b} \rightarrow J/\psi K^{-}p$~\cite{LHCbPc2015}. Four years later, the LHCb Collaboration updated their results, a new states $P_{c}(4312)$ was proposed, and
the $P_{c}(4450)$ was split to $P_{c}(4440)$ and $P_{c}(4457)$ states~\cite{LHCbPc2019}. These observations of the hidden-charm pentaquarks bring great interest in theoretical investigations. Excellent review on the hidden-charm pentaquarks can be found in Refs.~\cite{ChenHX0,LiuYR0,YangG0}.

Accordingly, it is timely to study the existence of the fully heavy pentaquarks. A systematic study on the mass spectra of the $S-$wave fully heavy pentaquarks $QQQQ\bar{Q}$ has been performed in the framework of the chromomagnetic interaction (CMI) model~\cite{An:2020jix}. Besides, the fully heavy pentaquarks were also investigated by the QCD sum rule approach~\cite{Zhang:2020vpz,Wang:2021xao}.

The quantum chromodynamics (QCD) is the underlying theory of the strong interaction. However, it is difficult to study the structure of the hadrons and the hadron-hadron interaction directly because of the nonperturbative properties of QCD in the low energy region. Although lattice QCD has made impressive progresses on nucleon-nucleon interactions and multiquark systems~\cite{Ishii:2006ec,Alexandrou:2001ip,Okiharu:2004wy}, the QCD-inspired quark model is still the main approach to study the hadron-hadron interactions and multiquark states. A common approach is the chiral quark model (ChQM)~\cite{Salamanca1}, in which the constituent quarks interact with each other through colorless Goldstone bosons exchange in addition to the colorful one-gluon-exchange and confinement, and the chiral partner $\sigma$ meson-exchange
is introduced to obtain the immediate-range attraction of hadron-hadron interaction. An alternative approach is the quark delocalization color
screening model (QDCSM), which was developed in 1990s with the aim of explaining the similarities between nuclear and molecular
forces~\cite{Wang:1992wi}. Both of these two models have been successfully applied to the study of the properties of deuteron, nucleon-nucleon and hyperon-nucleon
interactions~\cite{ChenLZ,ChenM}. Recently, these two models were used to study the fully heavy tetraquarks~\cite{Jin:2020jfc} and dibaryons~\cite{Huang:2020bmb}. Therefore, it is interesting to extend these two models to the fully heavy pentaquarks.
	
This paper is organized as follows. First, both the chiral quark model and the quark delocalization color screening model are introduced briefly in Section II. The way we construct the wave functions is presented, too. Then, the numerical results and discussions are given in Section III. Finally, the summary is given in Section IV.
	
\section{MODEL AND WAVE FUNCTIONS}
In this work, we use two models, ChQM and QDCSM, to investigate the fully heavy pentaquark systems. In this sector, we will introduce these two models and the wave functions of the pentaquark systems.
	
\subsection{Chiral quark model (ChQM)}
The Salamanca version was chosen as the representative of the chiral quark models, because the Salamanca group's work covers the hadron spectra,
nucleon-nucleon interaction, and multiquark states. The model details can be found in Ref.~\cite{Salamanca1}. Here we only give the Hamiltonian:

	\begin{align}
		H=&\sum_{i=1}^5\left(m_i+\frac{p_i^2}{2m_i}\right)-T_{cm} +\sum_{j>i=1}^5\left(V^{CON}_{ij}+V^{OGE}_{ij} \right)
	\end{align}
where $T_{cm}$ is the kinetic energy of the center of mass; $V^{CON}_{ij}$ and $V^{OGE}_{ij}$ are the interactions of the confinement and the one-gluon-exchange, respectively. For the fully-heavy systems, there is no $\sigma-$exchange and the Goldstone boson exchange. The central part of $V^{CON}_{ij}$ and $V^{OGE}_{ij}$ are shown below:
	\begin{align}
		V_{i j}^{CON}= & -\lambda_{i}^{c} \cdot \lambda_{j}^{c}\left(a_{c} r_{i j}^{2}+V_{0}\right)
	\end{align}
	\begin{align}
		V_{i j}^{OGE}= &\frac{\alpha_{s}}{4} \lambda_{i}^{c} \cdot \lambda_{j}^{c}\left[\frac{1}{r_{i j}}-\frac{\pi}{2} \delta\left(\mathbf{r}_{i j}\right)\left(\frac{1}{m_{i}^{2}}+\frac{1}{m_{j}^{2}}\right.\right. \nonumber \\
		&\left.\left.+\frac{4 \sigma_{i} \cdot \sigma_{j}}{3 m_{i} m_{j}}\right)\right]
	\end{align}
where $\alpha_{s}$ is the quark-gluon coupling constant. In order to cover the wide energy range from light to strange to heavy quarks, an effective scale-dependent quark-gluon coupling $\alpha_{s}(u)$ was introduced~\cite{Vijande:2004he}:
	\begin{align}
		\alpha_{s}(u) =& \frac{\alpha_{0}}{\ln \left(\frac{u^{2}+u_{0}^{2}}{\Lambda_{0}^{2}}\right)}
	\end{align}
The other symbols in the above expressions have their usual meanings. All parameters, which are fixed by fitting to the masses of
baryons with light flavors and heavy flavors, are taken from our previous work Ref.~\cite{Huang:2019esu}.
	
\subsection{Quark delocalization color screening model (QDCSM)}
Generally, the Hamiltonian of QDCSM is almost the same as that of ChQM, but with two modifications~\cite{Wang:1992wi}. The one is that there is no $\sigma$-meson exchange in QDCSM, and another one is that the screened color confinement is used between quark pairs reside in different clusters, aimed to take into account the QCD effect which has not yet been included in the two-body confinement and effective one gluon exchange. Since there is no $\sigma$-meson exchange interaction between the heavy quarks (c or b), the only difference here is the confinement interaction. The confining potential in QDCSM was modified as follows:
	\begin{align}
		V_{i j}^{CON} =& \left\{\begin{array}{l}
			-\lambda_{i}^{c} \cdot \lambda_{j}^{c}\left(a_{c} r_{i j}^{2}+V_{0}\right), \mathrm{i}, \mathrm{j} \text { in the same cluster } \\
			-\lambda_{i}^{c} \cdot \lambda_{j}^{c} a_{c} \frac{1-e^{-\mu_{i j} r_{i j}^{2}}}{\mu_{i j}}, \text { otherwise }
		\end{array}\right.
	\end{align}
where $\mu_{ij}$ is the color screening parameter, which is determined by fitting the deuteron properties, nucleon-nucleon scattering phase shifts, and hyperon-nucleon scattering phase shifts, respectively, with $\mu_{uu}=0.45~$fm$^{-2}$, $\mu_{us}=0.19~$fm$^{-2}$ and $\mu_{ss}=0.08~$fm$^{-2}$, satisfying the relation, $\mu_{us}^{2}=\mu_{uu}\mu_{ss}$~\cite{ChenM}. When extending to the heavy quark case, there is no experimental
data available, so we take it as a adjustable parameter $\mu_{cc}=0.01 \sim 0.001~$fm$^{-2}$ and $\mu_{bb}=0.001 \sim 0.0001~$fm$^{-2}$. We find the results
are insensitive to the value of $\mu_{cc}$ and $\mu_{bb}$. So in the present work, we take $\mu_{cc}=0.01~$fm$^{-2}$ and $\mu_{bb}=0.001~$fm$^{-2}$.

The quark delocalization in QDCSM is realized by specifying the single particle orbital wave function of QDCSM as a linear combination of left and right Gaussians, the single particle orbital wave functions used in the ordinary quark cluster model,
	\begin{eqnarray}
		\psi_{\alpha}(\mathbf{s}_i ,\epsilon) & = & \left(
		\phi_{\alpha}(\mathbf{s}_i)
		+ \epsilon \phi_{\alpha}(-\mathbf{s}_i)\right) /N(\epsilon) \nonumber \\
		\psi_{\beta}(-\mathbf{s}_i ,\epsilon) & = &
		\left(\phi_{\beta}(-\mathbf{s}_i)
		+ \epsilon \phi_{\beta}(\mathbf{s}_i)\right) /N(\epsilon) \nonumber \\
		N(\epsilon) & = & \sqrt{1+\epsilon^2+2\epsilon e^{-s_i^2/4b^2}} \label{1q} \\
		\phi_{\alpha}(\mathbf{s}_i) & = & \left( \frac{1}{\pi b^2}
		\right)^{3/4}
		e^{-\frac{1}{2b^2} (\mathbf{r}_{\alpha} - \frac{2}{5}\mathbf{s}_i)^2} \nonumber \\
		\phi_{\beta}(-\mathbf{s}_i) & = & \left( \frac{1}{\pi b^2}
		\right)^{3/4}
		e^{-\frac{1}{2b^2} (\mathbf{r}_{\beta} + \frac{3}{5}\mathbf{s}_i)^2} \nonumber
	\end{eqnarray}
Here $\mathbf{s}_i$, $i=1,2,...,n$ are the generating coordinates, which are introduced to expand the relative motion wavefunction. The mixing parameter $\epsilon(\mathbf{s}_i)$ is not an adjusted one but determined variationally by the dynamics of the multi-quark system itself. In this way, the multi-quark system chooses its favorable configuration in the interacting process. This mechanism has been used to explain the cross-over transition between hadron phase and quark-gluon plasma phase~\cite{Xu:2007oam}.
	
\subsection{Wave functions}
In this work, the resonating group method (RGM)~\cite{RGM}, a well-established method for studying
a bound-state or a scattering problem, is employed to calculate the energy of the fully heavy systems. The wave
function of the pentaquark system has four components: orbital, color, flavor and spin. The whole wave function can be obtained by multiplying them together according to a certain coupling relation. Then, the wave function of the pentaquark system is of the form:
	\begin{equation}
		\Psi = {\cal A } \left[[\psi^{L}\chi^{\sigma}]_{JM}\chi^{f}\chi^{c}\right]
	\end{equation}
where $\psi^{L}$, $\chi^{\sigma}$, $\chi^{f}$, and $\chi^{c}$ are the orbital, the spin, the flavor and the color wave functions respectively. The symbol ${\cal A }$ is the anti-symmetrization operator. In this work, we investigate the fully heavy pentaquark ($QQQQ\bar{Q}$) within the baryon-meson structure, so the ${\cal A }$ is defined as
	\begin{equation}
		{\cal A } = 1-P_{14}-P_{24}-P_{34}
	\end{equation}
where 1, 2, 3 and 4 stand for the quarks, while 5 stands for the antiquark.
	
The orbital wave function is in the form of	
	\begin{align}
		\psi^{L} =& {\psi}_{1}(\boldsymbol{R}_{1}){\psi}_{2}(\boldsymbol{R}_{2})\chi_{L}(\boldsymbol{R})
	\end{align}
where $\boldsymbol{R}_{1}$ and $\boldsymbol{R}_{2}$ are the internal coordinates for the cluster 1 and cluster 2. $\boldsymbol{R} = \boldsymbol{R}_{1}-\boldsymbol{R}_{2}$ is the relative coordinate between the two clusters 1 and 2. The ${\psi}_{1}$ and ${\psi}_{2}$ are the internal cluster orbital wave functions of the clusters 1 and 2, and $\chi_{L}(\boldsymbol{R})$ is the relative motion wave function between two clusters, which is expanded by the gaussian bases
	\begin{align}
		\chi_{L}(\boldsymbol{R}) =& \frac{1}{\sqrt{4 \pi}}\left(\frac{6}{5 \pi b^{2}}\right)^{3 / 4} \sum_{i = 1}^{n} C_{i} \\ \nonumber    &\times\int \exp \left[-\frac{3}{5 b^{2}}\left(\boldsymbol{R}-\boldsymbol{S}_{i}\right)^{2}\right] Y^{L}\left(\hat{\boldsymbol{S}}_{i}\right) d \hat{\boldsymbol{S}}_{i}
	\end{align}
where $\boldsymbol{S}_{i}$ is called the generate coordinate, $n$ is the number of the gaussian bases, which is determined by the stability of the results. By doing this, the integro-differential equation of RGM can be reduced to an algebraic equation, generalized eigen-equation. Then we can obtain the energy of the system by solving this generalized eigen-equation. The details of solving the RGM equation can be found in Ref.~\cite{RGM}.

For the spin wave function, we first construct the spin wave function of the $q^{3}$ and $q\bar{q}$ clusters, and then the total spin wave function of the pentaquark system is obtained by coupling the spin wave functions of two clusters together.
The spin wave functions of the $q^{3}$ and $q\bar{q}$ clusters are
	\begin{align}
		\chi_{\frac{3}{2}, \frac{3}{2}}^{\sigma}(3) & = \alpha \alpha \alpha \nonumber \\
		\chi_{\frac{3}{2}, \frac{1}{2}}^{\sigma}(3) & = \frac{1}{\sqrt{3}}(\alpha \alpha \beta+\alpha \beta \alpha+\beta \alpha \alpha) \\
		\chi_{\frac{3}{2},-\frac{1}{2}}^{\sigma}(3) & = \frac{1}{\sqrt{3}}(\alpha \beta \beta+\beta \alpha \beta+\beta \beta \alpha) \nonumber \\
		\chi_{1,1}^{\sigma}(2) & = \alpha \alpha \nonumber \\
		\chi_{1,0}^{\sigma}(2) & = \frac{1}{\sqrt{2}}(\alpha \beta+\beta \alpha) \nonumber \\
		\chi_{1,-1}^{\sigma}(2) & = \beta \beta \\
		\chi_{0,0}^{\sigma}(2) & = \frac{1}{\sqrt{2}}(\alpha \beta-\beta \alpha) \nonumber
	\end{align}
For pentaquark system, the total spin quantum number can be 1/2, 3/2 or 5/2, so the wave function of each spin quantum number can be written as follows.
	\begin{align}
		\chi_{\frac{1}{2}, \frac{1}{2}}^{\sigma 1}(5) = & \sqrt{\frac{1}{6}} \chi_{\frac{3}{2},-\frac{1}{2}}^{\sigma}(3) \chi_{1,1}^{\sigma}(2)-\sqrt{\frac{1}{3}} \chi_{\frac{3}{2}, \frac{1}{2}}^{\sigma}(3) \chi_{1,0}^{\sigma}(2)\nonumber \\ \nonumber &+\sqrt{\frac{1}{2}} \chi_{\frac{3}{2}, \frac{3}{2}}^{\sigma}(3) \chi_{1,-1}^{\sigma}(2) \\
		\chi_{\frac{3}{2}, \frac{3}{2}}^{\sigma 1}(5) = & \sqrt{\frac{3}{5}} \chi_{\frac{3}{2}, \frac{3}{2}}^{\sigma}(3) \chi_{1,0}^{\sigma}(2)-\sqrt{\frac{2}{5}} \chi_{\frac{3}{2}, \frac{1}{2}}^{\sigma}(3) \chi_{1,1}^{\sigma}(2)\nonumber \\
		\chi_{\frac{3}{2}, \frac{3}{2}}^{\sigma 2}(5) = & \chi_{\frac{3}{2}, \frac{3}{2}}^{\sigma}(3) \chi_{0,0}^{\sigma}(2) \nonumber \\
		\chi_{\frac{5}{2}, \frac{5}{2}}^{\sigma_{1}}(5) = & \chi_{\frac{3}{2}, \frac{3}{2}}^{\sigma}(3) \chi_{1,1}^{\sigma}(2)
	\end{align}
		
The flavor wave functions for the fully heavy pentaquark system we investigate here are very simple, which are
	\begin{eqnarray}
		\chi_{00}^{f m 1}=cccc\bar{c}  \\
		\chi_{00}^{f m 2}=bbbb\bar{b}
	\end{eqnarray}

The construction of the color wave function is similar to the spin wave function. We first write down the color wave function of the $q^{3}$ and $q\bar{q}$ clusters, and then the color wave function of the pentaquark system is obtained by coupling the two clusters together. For the color-singlet channel (two clusters are color-singlet), the color wave function is
	\begin{align}
	\chi^{c0} =& \frac{1}{\sqrt{6}}(r g b-r b g+g b r-g r b+b r g-b g r)\nonumber \\
	&\times\frac{1}{\sqrt{3}}(r \bar{r}+g \bar{g}+b \bar{b})
\end{align}
For the hidden-color channel (two clusters are color-octet), the symmetry of the color-octet cluster is $[21]$. The color wave function of the color-octet $q^{3}$ cluster can be written as

	\begin{align}
	\chi_{b1}^{c1}=&\frac{1}{\sqrt{6}}(2 r r g-r g r-g r r)\nonumber \\
	\chi_{b2}^{c1}=&\frac{1}{\sqrt{6}}(2 r r b-r b r-b r r) \nonumber \\
	\chi_{b3}^{c1}=&\frac{1}{\sqrt{6}}(r g g+g r g-2 g g r)    \nonumber \\
	\chi_{b4}^{c1}=&\frac{1}{\sqrt{12}}(2 r g b-r b g+2 g r b-g b r-b r g-b g r) \nonumber \\
	\chi_{b5}^{c1}=&\frac{1}{\sqrt{4}}(r b g-g b r+b r g-b g r) \nonumber \\
	\chi_{b6}^{c1}=&\frac{1}{\sqrt{6}}(r b b+b r b-2 b b r)\nonumber \\
	\chi_{b7}^{c1}=&\frac{1}{\sqrt{6}}(2 g g b-g b g- b g g) \nonumber \\
	\chi_{b8}^{c1}=&\frac{1}{\sqrt{6}}(g b b+b g b-2 b b g) \\
	\nonumber \\
	\chi_{b1}^{c2}=&\frac{1}{\sqrt{2}}(r g r-g r r) \nonumber \\
	\chi_{b2}^{c2}=&\frac{1}{\sqrt{2}}(r b r-b r r) \nonumber \\
	\chi_{b3}^{c2}=&\frac{1}{\sqrt{2}}(r g g-g r g)    \nonumber \\
	\chi_{b4}^{c2}=&\frac{1}{\sqrt{4}}(r b g+g b r-b r g-b g r) \nonumber \\
	\chi_{b5}^{c2}=&\frac{1}{\sqrt{12}}(2 r g b+r b g-2 g r b-g b r-b r g+b g r) \nonumber \\
	\chi_{b6}^{c2}=&\frac{1}{\sqrt{2}}(r b b-b r b) \nonumber \\
	\chi_{b7}^{c2}=&\frac{1}{\sqrt{2}}(g b b-b g b)  \nonumber  \\
	\chi_{b8}^{c2}=&\frac{1}{\sqrt{2}}(g b b-b g b)
	\end{align}
where superscript $c1$ and $c2$ represent the color symmetry of the quark 1 and 2 is symmetric and antisymmetric, respectively. The color wave function of the color-octet $q\bar{q}$ cluster can be written as
\begin{align}
\chi_{m1}^{c}=&r\bar{b},~~~~~~\chi_{m2}^{c}=-r\bar{g},~~~~\chi_{m3}^{c}=g\bar{b} \nonumber \\
\chi_{m4}^{c}=&\frac{1}{\sqrt{2}}(r\bar{r}-g\bar{g}),~~~~ \chi_{m5}^{c}=\frac{1}{\sqrt{6}}(2b\bar{b}-r\bar{r}-g\bar{g}) \nonumber \\
\chi_{m6}^{c}=&-b\bar{g},~~~~\chi_{m7}^{c}=g\bar{r},~~~~ \chi_{m8}^{c}=b\bar{r}
\end{align}
Then, the total color wave functions for the hidden-color channel is obtained by coupling the wave functions of two color-octet clusters according to the Clebsch-Gordan (CG) coefficients.
\begin{align}
	\chi^{c1}=\frac{1}{\sqrt{8}}(&\chi_{b1}^{c1}\chi_{m8}^{c}-\chi_{b3}^{c1}\chi_{m6}^{c}-\chi_{b2}^{c1}\chi_{m7}^{c}+\chi_{b4}^{c1}\chi_{m4}^{c}   \nonumber\\
	&-\chi_{b7}^{c1}\chi_{m2}^{c}+\chi_{b5}^{c1}\chi_{m5}^{c}-\chi_{b6}^{c1}\chi_{m3}^{c}+\chi_{b8}^{c1}\chi_{m1}^{c}) \\
	\chi^{c2}=\frac{1}{\sqrt{8}}(&\chi_{b1}^{c2}\chi_{m8}^{c}-\chi_{b3}^{c2}\chi_{m6}^{c}-\chi_{b2}^{c2}\chi_{m7}^{c}+\chi_{b4}^{c2}\chi_{m4}^{c}   \nonumber\\
&-\chi_{b7}^{c2}\chi_{m2}^{c}+\chi_{b5}^{c2}\chi_{m5}^{c}-\chi_{b6}^{c2}\chi_{m3}^{c}+\chi_{b8}^{c2}\chi_{m1}^{c})
\end{align}
	
Finally, we can acquire the total wave function by combining the wave functions of the orbital, spin, flavor and color parts together according to the quantum number of the pentaquark systems.

\section{The results and discussions}
In this work, we investigate the $S-$wave fully heavy pentaquark systems $cccc\bar{c}$ and $bbbb\bar{b}$ in the framework of two quark models, the ChQM and the QDCSM. The quantum numbers of these systems are $J^P = 1/2^-$, $3/2^-$, and $5/2^-$.

To find out if there is any bound state in such fully heavy pentaquark systems, we do a dynamic bound-state calculation here.
The single-channel calculation, as well as the channel-coupling are carried out. All the calculation results for $cccc\bar{c}$ and $bbbb\bar{b}$ systems in two models are listed in Tables~\ref{ccccc} and \ref{bbbbb}, respectively. The second column headed with $channel$ denotes the physical contents of each channel; $E_{th}$ and $E_{sc}$ represent the threshold and the energy of each channel. As for the result of the channel-coupling calculation, $E_{cc1}$ stands for the coupling of only color-singlet channels, and $E_{cc2}$ stands for the coupling of both the color-singlet channels and the hidden-color channels. Besides, the subscript $8$ is marked to represent the color-octet channels, and the asterisk '$*$' is added to the energy, which is lower than the threshold of the corresponding channel.

\begin{table*}[ht]
		\caption{\label{ccccc}The energies of the $cccc\bar{c}$ pentaquark systems (in MeV).}
		\begin{tabular}{c c c c c c c c c c c} \hline\hline
			~  & ~  & \multicolumn{4}{c}{ChQM} && \multicolumn{4}{c}{QDCSM}   \\
			\cline{3-6} \cline{8-11}
			$J^P$  &channel & $E_{th}$ & $E_{sc}$ & $E_{cc1}$ & $E_{cc2}$ &~& $E_{th}$& $E_{sc}$& $E_{cc1}$& $E_{cc2}$\\ \hline
			$\frac{1}{2}^-$ &$\Omega_{ccc}J/\psi$ & ~8427.7~ & ~8428.9~ & ~ & ~8427.5* && ~8537.4~ & ~8538.6~ & ~& ~8536.4*\\
			~ &$\Omega_{ccc8}\eta_{c8}$& ~ & 8658.4 & ~ & ~ && ~ & ~8786.8~ & ~ & ~ \\
			~ &$\Omega_{ccc8}J/\psi_8$  & ~ & 8739.8 & ~ & ~ && ~ & ~8897.5~ & ~ & ~ \\
			$\frac{3}{2}^-$&$\Omega_{ccc}\eta_{c}$ & 8424.9 & 8426.8 & 8425.6 & 8425.4&& 8534.8& 8536.8& 8535.5 &8535.0 \\
			~ & $\Omega_{ccc}J/\psi$ & 8427.7 & 8429.5 & ~ & ~ && 8537.4 & 8539.2 & ~ & ~    \\
			~ & $\Omega_{ccc8}J/\psi_8$ & ~   & 8616.3 & ~ & ~ && ~ &8728.1 & ~ & ~  \\
			$\frac{5}{2}^- $ & $\Omega_{ccc}J/\psi$ & 8427.7 & 8429.8 & ~ && ~ & 8537.4 & 8539.4 & ~ & ~  \\  \hline\hline
		\end{tabular}
	\end{table*}
	
	\begin{table*}[ht]
		\caption{\label{bbbbb}The energies of the $bbbb\bar{b}$ pentaquark systems (in MeV).}
		\begin{tabular}{c c c c c c c c c c c} \hline\hline
			~  & ~  & \multicolumn{4}{c}{ChQM} && \multicolumn{4}{c}{QDCSM}   \\
			\cline{3-6} \cline{8-11}
			$J^P$  &channel & $E_{th}$ & $E_{sc}$ & $E_{cc1}$ & $E_{cc2}$ &~& $E_{th}$& $E_{sc}$& $E_{cc1}$& $E_{cc2}$\\ \hline
			$\frac{1}{2}^-$&$\Omega_{bbb}\Upsilon$ & ~25178.9~ & ~25179.4~ & ~ & ~25166.4* && ~25275.2~ & ~25275.6~ & ~& ~25259.0*\\
			~ &$\Omega_{bbb8}\eta_{b8}$& ~ & 25308.7 & ~ & ~ && ~ & ~25388.6~ & ~ & ~ \\
			~ &$\Omega_{bbb8}\Upsilon_8$  & ~ & 25380.1 & ~ & ~ && ~ & ~25496.0~ & ~ & ~ \\
			$\frac{3}{2}^-$&$\Omega_{bbb}\eta_{b}$ & 25178.7 & 25179.4 & 25179.2 &~25166.3* && 25275.0 & 25275.7& 25275.5 &~25258.9* \\
			~ & $\Omega_{bbb}\Upsilon$ & 25178.9 & 25179.6 & ~ & ~ && 25275.2 & 25275.5 & ~ & ~    \\
			~ & $\Omega_{bbb8}\Upsilon_8$ & ~   & 25272.8 & ~ & ~ && ~ &25335.3 & ~ & ~  \\
			$\frac{5}{2}^-$& $\Omega_{bbb}\Upsilon$ & 25178.9 & 25179.2 & ~ && ~ & 25275.2 & 25275.9 & ~ & ~  \\  \hline\hline
		\end{tabular}
	\end{table*}

\subsection{Fully charm pentaqurks}
The energies of the fully charm pentaquark systems are listed in Table~\ref{ccccc}, including both the single channel and channel-coupling results. Obviously, the energies of every single channel are above the corresponding theoretical threshold in both two models, which means that each color-singlet channel is unbound. The energy of each hidden-color channel is much higher than that of the color-singlet channel. After the channel-coupling calculation, the energy of the $J^P=1/2^-$ state is lower than the threshold of $\Omega_{ccc} J/\psi$, the binding energy of which is $-0.2$ MeV in ChQM and $-1.0$ MeV in QDCSM. However, for the $J^P=3/2^-$ state, the energy is still higher than the theoretical threshold of $\Omega_{ccc} \eta_{c}$, which indicates that there is no any bound state lower than $\Omega_{ccc} \eta_{c}$ for the $J^P=3/2^-$ system in both two models. Besides, there is only one channel for the $J^P=5/2^-$ state, and it is not bound either.

As we mentioned in Section II that the parameters we used in this work are taken from our previous work of Ref.~\cite{Huang:2019esu}, where the masses of the fully heavy baryon $\Omega_{ccc}$ and $\Omega_{bbb}$ are respectively $5068.8$ MeV and $15111.6$ MeV in ChQM, and $5068.8$ MeV and $15111.6$ MeV in QDCSM. When extending to the mesons, none of the parameters were readjusted. So the masses of the mesons $\eta_{c}$, $J/\psi$, $\eta_{b}$ and $\Upsilon$ are not well reproduced here. They are respectively $3355.8$ MeV, $3358.6$ MeV, $10067.1$ MeV and $10067.3$ MeV in ChQM, and $3400.2$ MeV, $3402.8$ MeV, $10105.7$ MeV and $10105.9$ MeV in QDCSM.
In order to minimize the theoretical errors and to compare calculated results to the experimental data in future, we shift the mass of the pentaquark system here.
Generally, the mass of a molecular pentaquark can be written as $M^{the.}=M^{the.}_{1}+M^{the.}_{2}+B$, where $M^{the.}_{1}$ and $M^{the.}_{2}$ stand for the theoretical masses of a baryon and a meson respectively, and $B$ is the binding energy of this molecular state.
To minimize the theoretical errors, we can shift the mass of molecular pentaquark to $M=M^{exp.}_{1}+M^{exp.}_{2}+B$, where the experimental
values of a baryon and a meson are used. Since there is no experimental value of the fully heavy $\Omega_{ccc}$ and $\Omega_{bbb}$, we use the value predicted by the lattice QCD calculation~\cite{Brown:2014ena}, which is $4796$ MeV and $14366$ MeV, respectively.
Taking the fully charm state $\Omega_{ccc} J/\psi$ with $J^P=1/2^-$ in ChQM as an example, the calculated mass of pentaquark is
$8427.5$ MeV, then the binding energy $B$ is obtained by subtracting the theoretical masses of
$\Omega_{ccc}$ and $J/\psi$, $8427.5-5069.1-3358.6=-0.2$ (MeV). Using the Lattice QCD mass of $\Omega_{ccc}$ and the experimental mass of $J/\psi$, the mass of this pentaquark is $M=4796+3096.9+(-0.2)=7892.7$ (MeV). The same approach is used in the QDCSM, and the mass of this pentaquark $7891.9$ MeV is arrived. So we finally obtain a fully charm pentaquark state with $J^P=1/2^-$, the mass of which is $7891.9 \sim 7892.7$ MeV.

In addition, to confirm the existence of the fully charm pentaquark state with $J^P=1/2^-$, the spacial distribution feature of this state is investigated. The relative wave function between $\Omega_{ccc}$ and $J/\psi$ as a function of their distance $r$ in both two models is plotted in Fig. 1. To compare the result of the single channel and that of the channel-coupling, both of them are shown in Fig. 1. It is clearly that the behavior of the relative wave function of the single channel stands for the unbound nature of the single $\Omega_{ccc} J/\psi$, while the one of the channel-coupling calculation indicates the bound property of this state. Therefore, the effect of the channel-coupling is crucial for forming a bound state for the fully charm pentaquark system with $J^P=1/2^-$.

\begin{figure}[h]
		\includegraphics[width=8cm]{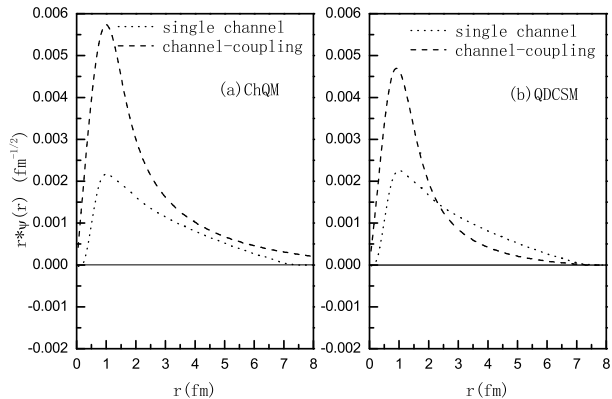}\
		\caption{\label{wave function}The relative wave function of the fully charm pentaquark system with $J^P=1/2^-$.}
\end{figure}

To understand the interaction between two heavy hadrons, we continue to study the effective potentials, which are shown in Figs. 2-4. The effective potential between two colorless clusters is defined as $V(S)=E(S)-E(\infty)$, where $E(S)$ is the diagonal matrix element of the Hamiltonian of the system in the generating coordinate.
From Figs. 2-4 we find that the results in both ChQM and QDCSM are similar. For the $J^P=1/2^-$ system (Fig. 2), one sees that the potential for the color-singlet channel $\Omega_{ccc} J/\psi$ is attractive. However, the attraction is not strong enough to form a bound state. So this single channel is unbound in the bound-state calculation. It is obvious that the attraction becomes larger by the channel-coupling calculation, which leads to the bound state. For the $J^P=3/2^-$ system (Fig. 3), we can see that the effective potentials for the $\Omega_{ccc} \eta_{c}$ and $\Omega_{ccc} J/\psi$ are repulsive, so no bound states can be formed in these two channels. Although the potential is attractive by the channel-coupling, it is not strong enough to form a bound state. So we do not obtain the bound state for the $J^P=3/2^-$ system. For the $J^P=5/2^-$ system (Fig. 4), the potential for the $\Omega_{ccc} J/\psi$ is clearly repulsive. Besides, the repulsion increases greatly when the two hadrons $\Omega_{ccc}$ and $J/\psi$ get very close. This is due to the Pauli exclusion principle. Considering that the four $c$ quarks in the pentaquark system of $J^P=5/2^-$ are identical particles, they have the same state in the orbital, flavor, and spin wave functions. But there are only three color states, which are red, green and blue. Because of this, these four quarks are difficult to stay together. That's why we cannot obtain the $cccc\bar{c}$ pentaquark with $J^P=5/2^-$.

Besides, we also find the results in both two models are almost the same, this is because that the quarks are too heavy to run, resulting in the value of the quark delocalization parameter $\epsilon$ in QDCSM is close to 0. The color screening parameter in QDCSM is also very small because of the heavy quarks, which makes the difference of the confinement between two models is very small. So both the effect of the quark delocalization and the color screening in QDCSM is very small
in such fully heavy system. Meanwhile, the $\sigma$ meson exchange is also inoperative in ChQM, which make the coincident
results of two models.

	\begin{figure}[h]
		\centering
		\includegraphics[width=8cm]{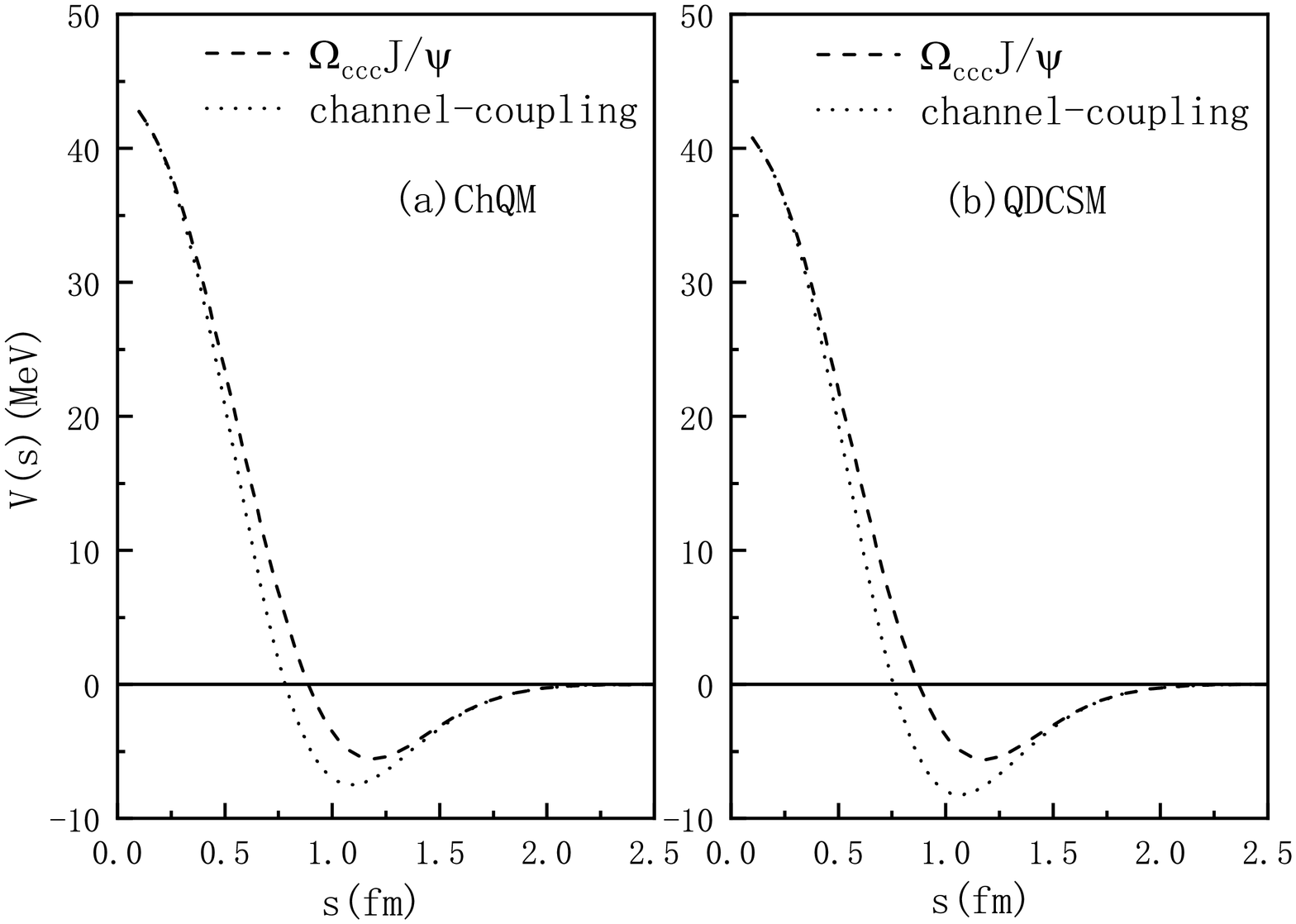}\
		\caption{\label{c 0.5 coupling}The effective potentials of the fully charm system with $J^P=1/2^-$ in ChQM and QDCSM.}
	\end{figure}

	\begin{figure}[h]
		\centering
		\includegraphics[width=8cm]{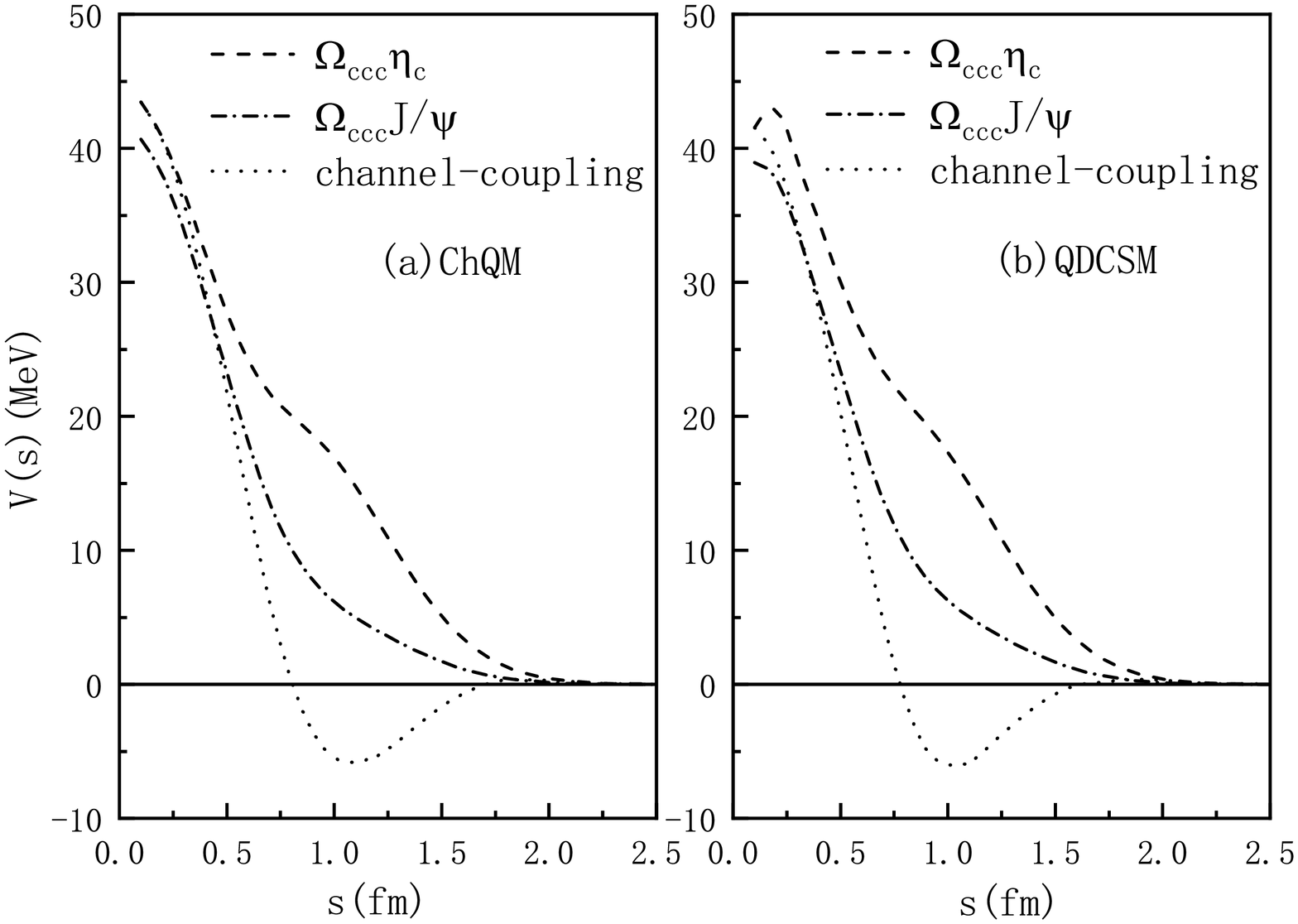}\
		\caption{\label{c 1.5 coupling}The effective potentials of the fully charm system with $J^P=3/2^-$ in ChQM and QDCSM.}
	\end{figure}
	\begin{figure}[h]
		\centering
		\includegraphics[width=8cm]{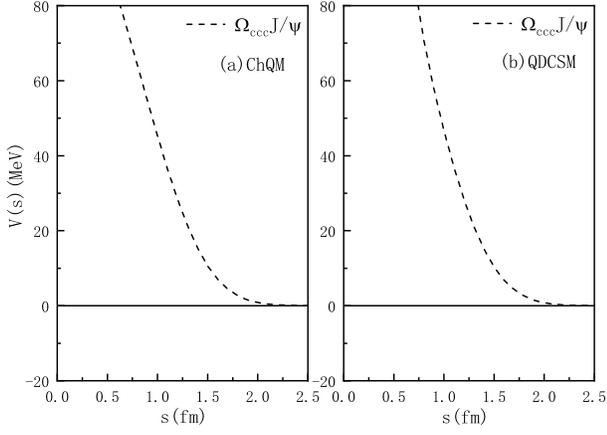}\
		\caption{\label{c 2.5}The effective potentials of the fully charm system with $J^P=5/2^-$ in ChQM and QDCSM.}
	\end{figure}

\subsection{Fully bottom pentaqurks}
In the previous discussion, the fully charm pentaquarks were investigated. We also extend the study to the fully bottom pentaquarks because of the heavy flavor
symmetry. The numerical results are listed in Table~\ref{bbbbb}. We find that the results are similar to the fully charm pentaquarks. All the color-singlet channel is unbound, and the energy of the hidden-color channel is much higher than the color-singlet channel. After the channel-coupling calculation, a bound state is obtained for the $J^P=1/2^-$ system, with the binding energy of $-12.5$ MeV in ChQM and $-16.2$ MeV in QDCSM. By performing the mass shift, we finally obtain the mass of this pentaquark, which is $23810.1 \sim 23813.8$ MeV. Besides, the state with $J^P=3/2^-$ is also bound with the help of channel-coupling. The binding energy is $-12.4$ MeV in ChQM and $-16.1$ MeV in QDCSM. By shifting the mass, the mass of the fully bottom pentaquark state with $J^P=3/2^-$ is $23748.2 \sim 23752.3$ MeV.

Additionally, the effective potentials of the fully bottom pentaquark systems in both ChQM and QDCSM are also studied, which are shown in Figs. 5-7. Obviously, the results are also similar to the fully charm pentaquark systems.
	
	\begin{figure}[h]
		\includegraphics[width=8cm]{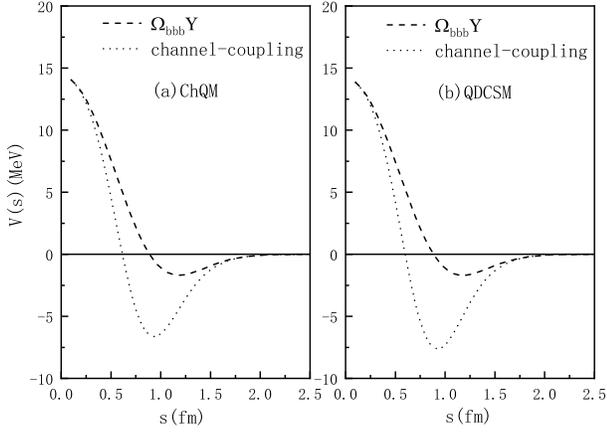}\
		\caption{\label{b 0.5 coupling}The effective potentials of the fully bottom system with $J^P=1/2^-$ in ChQM and QDCSM.}
	\end{figure}
	\begin{figure}[h]
	\centering
	\includegraphics[width=8cm]{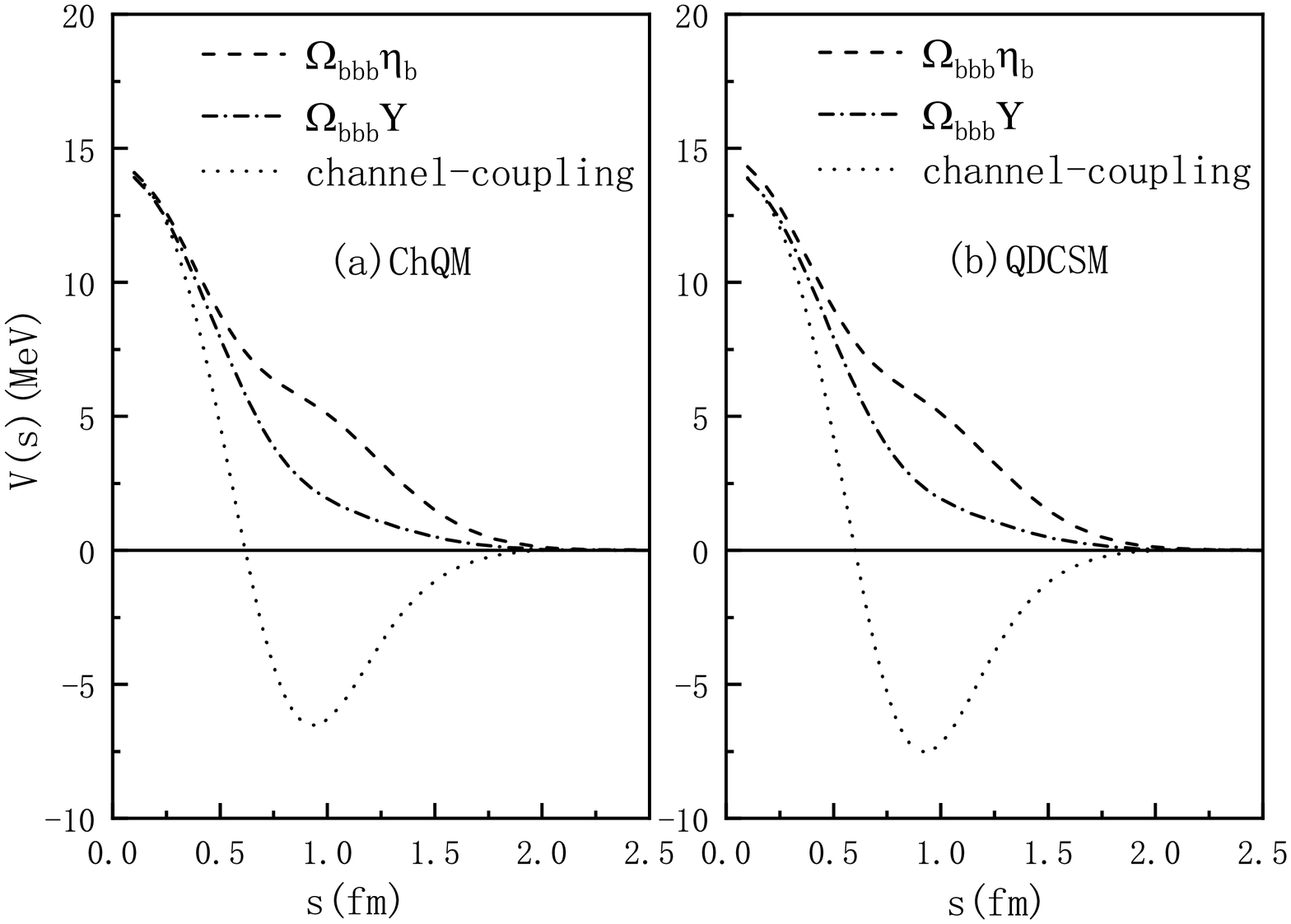}\
	\caption{\label{b 1.5 coupling}The effective potentials of the fully bottom system with $J^P=3/2^-$ in ChQM and QDCSM.}
\end{figure}
\begin{figure}[h]
	\centering
	\includegraphics[width=8cm]{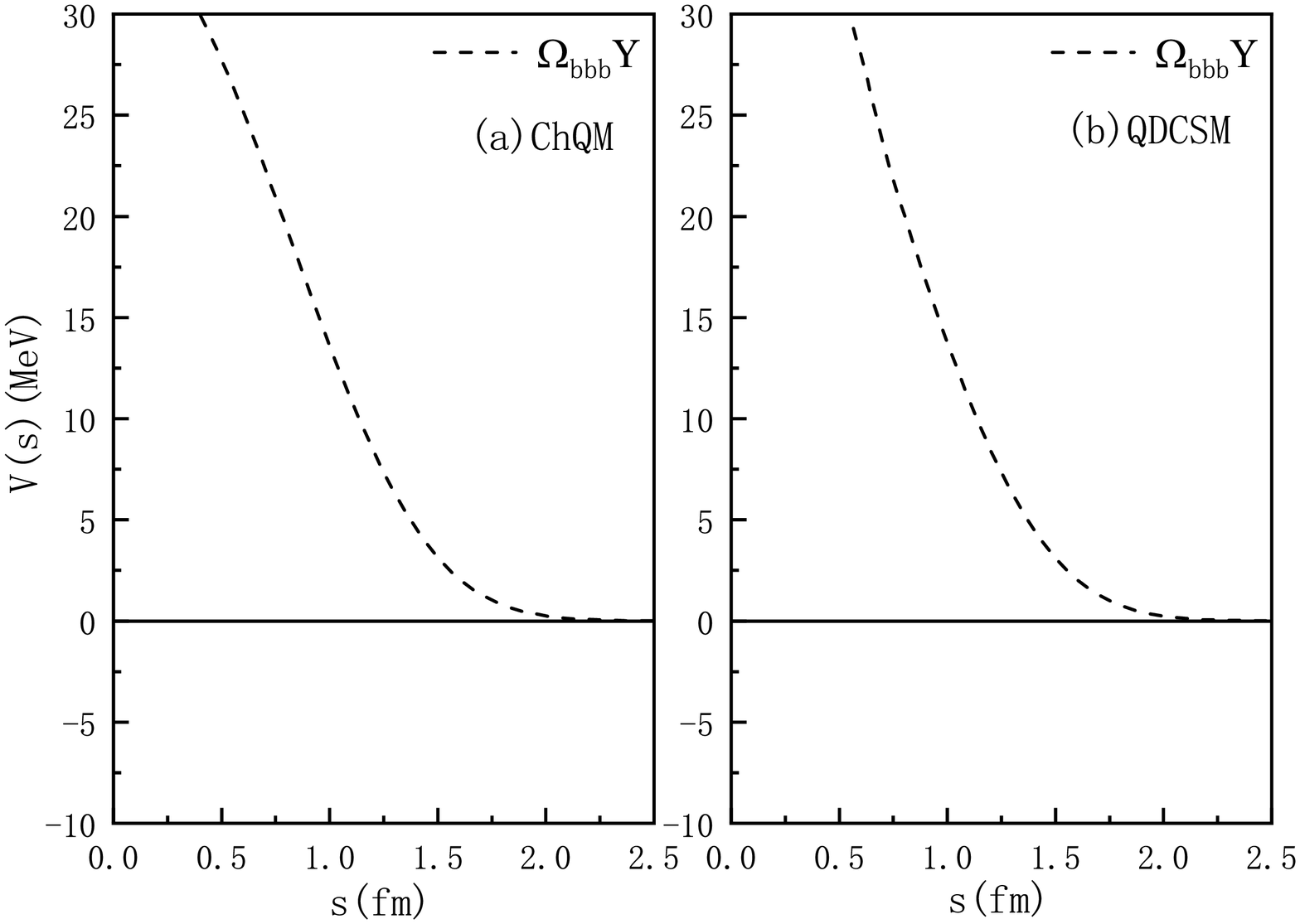}\
	\caption{\label{b 2.5}The effective potentials of the fully bottom system with $J^P=5/2^-$ in ChQM and QDCSM.}
\end{figure}
	
\section{Summary}
In this work, we systematically investigate the low-lying fully heavy pentaquark systems $cccc\bar{c}$ and $bbbb\bar{b}$ in two quark models: ChQM and QDCSM.
The dynamic bound-state calculation is carried out to search for any bound state in the fully heavy systems. Both the single channel and the channel coupling calculation are performed to explore the effect of the multi-channel coupling. Meanwhile, the relative wave function of the fully charm pentaquark state is studied to confirm the existence of the bound state. Besides, an adiabatic calculation of the effective potentials is added to explore the interaction between the heavy baryon and the heavy meson.

The numerical results show that the effect of the channel-coupling is important for forming a bound state of the fully heavy pentaquark system and the conclusions are consistent in both ChQM and QDCSM. With the help of the channel-coupling, we obtain three fully heavy pentaquarks, which are the fully charm pentaquark with $J^P = 1/2^-$ and the mass of $7891.9 \sim 7892.7$ MeV, the fully bottom pentaquark with $J^P=1/2^-$ and the mass of $23810.1 \sim 23813.8$ MeV, and the fully bottom pentaquark with $J^P = 3/2^-$ and the mass of $23748.2 \sim 23752.3$ MeV. All these fully heavy pentaquark states are worth looking for in future experiments.

In Ref.~\cite{An:2020jix}, the study of the fully heavy pentaquarks in the CMI model found two $cccc\bar{c}$ states with $J^P=1/2^-$ and $3/2^-$ respectively, but the masses of them were above the thresholds of the $\Omega_{ccc} J/\psi$ and $\Omega_{ccc} \eta_{c}$ respectively. The channel-coupling may bring down the masses of these states. In the framework of QCD sum rules, both the $cccc\bar{c}$ state and $bbbb\bar{b}$ state are predicted in two works~\cite{Zhang:2020vpz} and ~\cite{Wang:2021xao}. However, the masses and the structure of these states are different in two works. More theoretical studies and experimental efforts, which may shed more light on the nature of the fully heavy pentaquark states, are expected.

\acknowledgments{This work is supported partly by the National Science Foundation
of China under Contract Nos. 11675080, 11775118 and 11535005.}

\end{document}